\documentclass[aip,jap,amsmath,amssymb,groupedaddress,reprint]{revtex4-1}
\usepackage{graphicx}
\usepackage{dcolumn}
\usepackage{bm}
\usepackage[utf8]{inputenc}
\usepackage[T1]{fontenc}
\usepackage{mathptmx}
\usepackage{etoolbox}
\usepackage{color}
\usepackage{multirow}
\makeatletter
\def\@email#1#2{%
 \endgroup
 \patchcmd{\titleblock@produce}
  {\frontmatter@RRAPformat}
  {\frontmatter@RRAPformat{\produce@RRAP{*#1\href{mailto:#2}{#2}}}\frontmatter@RRAPformat}
  {}{}
}%
\makeatother
\begin{document}

\preprint{AIP/123-QED}

\title{Spectral Recognition of Magnetic Nanoparticles with Artificial Neural Networks}

\author{David Slay}
\author{Michalis Charilaou}
\email[michalis.charilaou@louisiana.edu]{}
\affiliation{Department of Physics, University of Louisiana at Lafayette, Lafayette, Louisiana 70504}

\date{\today}

\begin{abstract}
Ferromagnetic resonance (FMR) spectroscopy is a powerful method for quantifying internal magnetic anisotropy fields in nanoparticles, which is important in a wide range of biomedical and storage applications. The interpretation of FMR spectra, however, can only be achieved with the use of an appropriate model, and no inverse methods are available to extract internal fields from FMR spectra. Here, we present the use of artificial neural networks for spectral recognition, i.e., to identify the internal magnetic anisotropy field from the FMR spectrum. We trained two different types of networks, a convolutional neural network and a multi-layer perceptron, by feeding the networks pre-computed FMR spectra labeled with the corresponding anisotropy fields. Testing of the trained networks with unseen spectra showed that they successfully predict the correct anisotropy fields and, surprisingly, the networks performed well for data that was beyond their training range. These results show the promise of using artificial neural networks for accelerated high-throughput analysis of magnetic materials and nanostructures; for example they could serve in automatizing and optimizing exploration missions where nanomagnetic signals are often used as proxies.
\end{abstract}




\maketitle

\section{Introduction}

The ability to quantify the magnetic properties of nanoparticles and their assemblies is essential in enabling the progress of numerous applications, particularly in biomedicine \cite{Pankhurst2003,mcbain2008,colombo2012,duerr2013,perigo2015,chantrell2015,conolly2018,materon2021} and bio-geological exploration \cite{weiss2004,gehring2011,blattmann2020}. Specifically, it is important to understand the association between internal anisotropy fields and inter-particle interaction fields with the performance and overall response of the nanoparticles to external stimuli \cite{myrovali2016,evans2020,slay2021}, and this requires techniques that enable the quantification of internal magnetic fields on the nanoscale. One of the most powerful techniques for the investigation of magnetic nanoparticles is ferromagnetic resonance spectroscopy (FMR) \cite{kittel1948}, because it can provide selective and quantitative information about the magnetic anisotropy fields. The interpretation of FMR spectra, however, requires the appropriate theoretical model in order to link the material parameters, i.e., the internal anisotropy fields, to the shape of the FMR spectrum \cite{kittel1948,vonsovskii1966,vonBardeleben2020}. Here, we show that artificial neural networks can be trained to identify material parameters from the FMR spectrum of powder samples, thus providing a novel way for high throughput spectral recognition of magnetic nanoparticles.

In an FMR experiment, a bias field $B$ is applied to magnetize the sample and a microwave field $B_\mathrm{mw}$, with a frequency in the range 10 -- 100 GHz, is applied orthogonally to the bias field. Typically, the frequency of the microwave is fixed and the spectrum is recorded, while sweeping the bias field, by measuring the out-of-phase susceptibility $\chi''$, which is a measure of dissipation, i.e., how much of the microwave energy the sample absorbs. With increasing strength of $B$, the Larmor precession frequency of the magnetic moments in the sample increases, and when the precession frequency matches the frequency of the microwave field, resonance occurs and the sample absorbs microwave radiation at the maximum. In ferromagnetic materials, the Larmor frequency depends on the effective field in the sample, which consists of both external and internal fields, and the resonance field $B_\mathrm{res}$ therefore contains information about the internal fields in the material: if the internal anisotropy field is in the same direction as the bias field, a smaller bias field is needed to accomplish resonance, and reversely, if the anisotropy field works against the bias field, a higher bias field is needed for resonance. Given that, one can quantify the magnetic anisotropy of single ferromagnetic crystals by measuring the resonance field as a function of the crystal orientation. In powder samples, e.g. ensembles of randomly oriented nanoparticles, one records the resonances at each orientation with a single measurement, and the FMR spectrum is a convolution of multiple resonances. Numerous FMR investigations of magnetic nanoparticles have provided insight to the modeling of FMR spectra using forward modeling \cite{debiasi1978,berger1999,berger2000,berger2001,sukhov2008,debiasi2003,debiasi2013,raikher1994,raikher2016,noginova2007,noginova2008,stepanov1999,schmool2000,li2009,usselman2012}, for example by solving the Landau-Lifshitz-Gilbert (LLG) equation of motion \cite{berger1999,berger2000,berger2001,sukhov2008} or by solving the athermal resonance equation \cite{smit1955,suhl1955,beselgia1988}. Despite having a solid foundation for calculating FMR spectra, the direct comparison between theory and experiment remains challenging because it constitutes an inverse problem. Recent advances in computational methods can provide solutions to this problem. For example, the use a genetic algorithm to decode FMR spectra of nanoparticle assemblies was shown to be promising, but for a limited range of systems \cite{usov2022}. Artificial neural networks (NN) \cite{rosenblatt1962,hilton2015} have recently been rapidly advancing computational techniques with very promising performances. In magnetism, NN have been shown to predict the response of the magnetization to an external field \cite{schrefl2019}, to reproduce hysteresis based on the Preisach model \cite{makaveev2001,antonio2021} and based on experimental data \cite{yekta2019}, and they also showed promise in solving inverse problems in magnetostatics and micromagnetics \cite{schrefl2022}. 

Here, we show that artificial neural networks can determine anisotropy fields in magnetic nanoparticles by associating FMR spectra with material parameters, given the appropriate training with data produced by a Stoner-Wohlfarth-type model \cite{charilaou2011,charilaou2017}. Below, we discuss the theoretical model of single-domain magnetic nanoparticles with uniaxial symmetry, which is a good approximation for a wide range of real systems. Based on the model, FMR spectra were computed for a wide range of material parameters, and the computed data were used to train two different architectures of artificial NN: a Convolution Neural Network (CNN) and a Multi-Layer Perceptron (MLP). The NN were tested by providing them with unseen FMR spectra and comparing the predicted material parameters to the ground truth, and both NN successfully recovered the anisotropy field of magnetic nanoparticles from their FMR spectrum. Strikingly, both NN were found to perform well even with FMR data that were beyond their training set. This highlights that machine learning techniques can be used to overcome bottlenecks of data analysis in magnetism and provide high throughput solutions to inverse problems in analyzing material properties in systems that are vital for a wide range of technologies and applications.


\section{Theory}
In our model, we consider single-domain particles with azimuthal symmetry, i.e., a uniaxial easy anisotropy axis. The magnetization and field vectors are parametrized by $\mathbf{M}=M(\theta)$ and $\mathbf{B}=B(\vartheta)$, respectively. The total energy density for a single particle is 
\begin{align}
E &=  -  \frac{1}{2}M_\mathrm{s}B_\mathrm{u} \cos^2 \theta - M_\mathrm{s}B \cos\left(\theta-\vartheta\right) 
\end{align}
where $B_\mathrm{u}$ is the uniaxial anisotropy field. This field can be either due to the crystal structure or the shape of the particle \cite{guimaraes2009}, but in both cases it has the same $\theta$ dependence and is therefore indistinguishable \cite{charilaou2017}. The details of $B_\mathrm{u}$ are beyond the scope of this work; here we focus on how to find $B_\mathrm{u}$ from the FMR spectrum.

In order to obtain the resonance condition, we calculate the angle-dependent resonance fields $B_\mathrm{res}(\theta)$ at fixed frequency $\omega$ from the resonance equation \cite{smit1955,suhl1955,beselgia1988}:
\begin{align}
\left( \frac {\omega} {\gamma} \right)^2 &= \frac {1} {M_\mathrm{s}^2}\partial_{\theta \theta}E \left( \frac {\cos \theta} {\sin \theta} \partial_{\theta}E \right),
\label{eq:res-eq}
\end{align}
\normalsize
\noindent where $\partial_\theta E$ and $\partial_{\theta\theta} E$ is the first and second derivatives of $E$ with respect to $\theta$, respectively, and $\gamma=g\mu_\mathrm{B}/\hbar$ is the gyromagnetic ratio with the Bohr magneton $\mu_\mathrm{B}$, the reduced Planck's constant $\hbar$ and the spectroscopic splitting factor $g$, which is the ratio of the total magnetic moment to the spin angular momentum \cite{morrish2001}, and for this study it was kept at $g=2.1$, which is a typical value for magnetite (Fe$_3$O$_4$) \cite{bickford1950}. With that, the resonance equation is
\begin{align}
\left( \frac {\omega} {\gamma} \right)^2 &= \left[ B \cos\left(\theta-\vartheta\right)+ B_\mathrm{u}\cos 2 \theta \right] \nonumber \\
&\times \left[ B \cos\left(\theta-\vartheta\right)+ B_\mathrm{u}\cos\theta^2 \right] \; 
\label{eq:res-full2}
\end{align}
and it has a solution at $\left|B\right|=B_\mathrm{res}(\vartheta)$.

\begin{figure}[h]
	\centering
		\includegraphics[width=1.00\columnwidth]{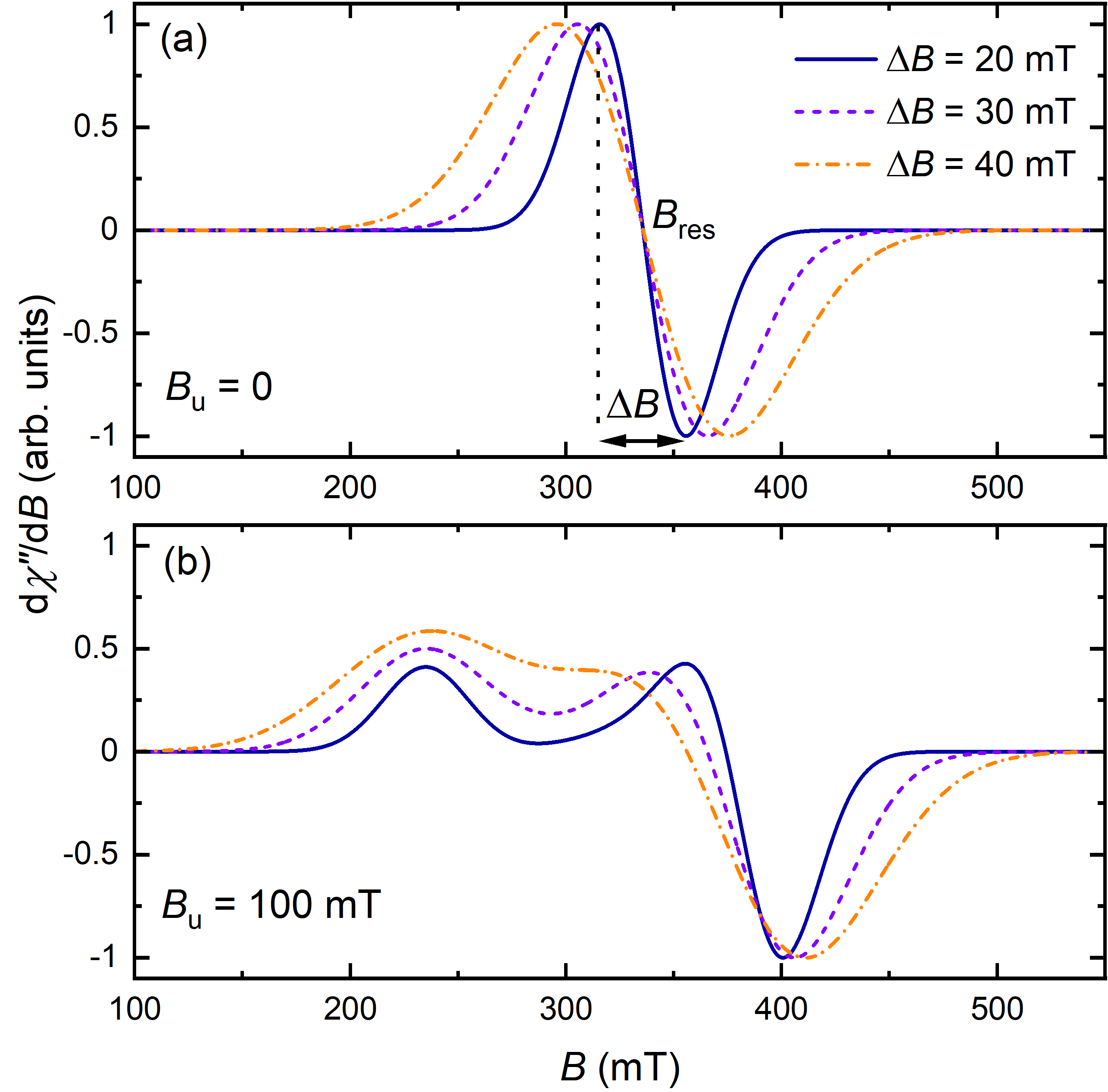}
	\caption{Examples of FMR spectra for (a) a population of isotropic single-domain particles with $B_0=0$ and (b) a population of anisotropic particles with $B_0=100$ mT with an even directional distribution, at different linewidths. In panel (a) the resonance field $B_\mathrm{res}$ and the linewidth $\Delta B$ are shown.}
	\label{FMR_Example}
\end{figure}

Considering that the magnetization direction $\theta$ depends on $\vartheta$ and the strength of $B$, Eq. \ref{eq:res-full2} needs to be solved at equilibrium, i.e., where $\partial_\theta E=0$. Given that, we compute the resonance field $B_\mathrm{res}(\vartheta)$ from Eq. \ref{eq:res-full2} and produce individual FMR spectra for all $\vartheta$. The individual FMR spectra empirically have the form of Gaussian derivatives, i.e., the derivative of the imaginary part of the susceptibility $\chi''$ is
\begin{equation}
\frac{d\chi''}{dB}= \sum_{\vartheta}p(\vartheta)\frac{2\left[B-B_\mathrm{res}(\vartheta)\right]}{\Delta B^3(\vartheta)\sqrt{2\pi}}e^{-\left[\frac{B-B_\mathrm{res}(\vartheta)}{\Delta B(\vartheta)}\right]^2}
\end{equation}
where $B_\mathrm{res}$ is the resonance field and depends on $B_\mathrm{u}$, $\Delta B$ is the individual linewidth, which depends on the magnetization damping mechanisms and the measurement frequency \cite{gilbert2004,Hickey2009} (see Fig. \ref{FMR_Example}a). The weight $p(\vartheta)=\sin\vartheta/4\pi$ projects the contribution of each resonance onto the field axis, where we assume a uniform orientational distribution of the particle axes.

Figure \ref{FMR_Example} shows examples of computed FMR spectra of populations with uniform orientational distribution, comparing isotropic particles ($B_\mathrm{u}=0$, Fig. \ref{FMR_Example}a) and anisotropic particles ($B_\mathrm{u}=100$ mT, Fig. \ref{FMR_Example}b), and illustrates the typical form of a resonance spectrum and how the symmetry of the spectrum changes with increasing $B_\mathrm{u}$.

\begin{figure}[h]
	\centering
		\includegraphics[width=1.00\columnwidth]{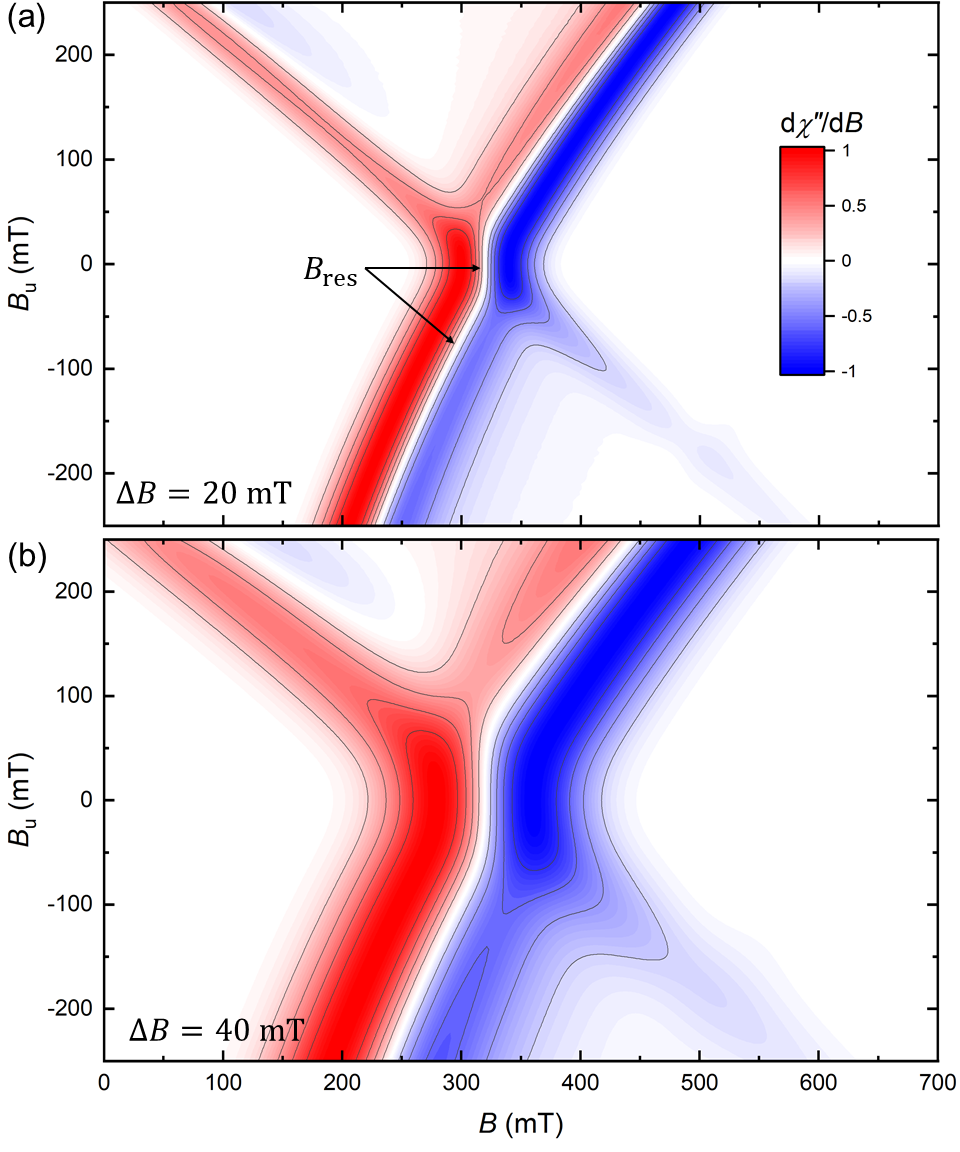}
	\caption{Training data consisting of FMR spectra for uniaxial anisotropy field in the range of $-250$ mT $\leq B_\mathrm{u}\leq +250$ mT and different linewidth of the convoluted spectrum. Here, only two examples with $\Delta B=20$ mT and 40 mT are shown. The contour color corresponds to the value of the derivative of the absorption $d\chi ''/dB$, and the main resonance field $B_\mathrm{res}$ can be traced by the white ridge as indicated in panel (a).}
	\label{contour}
\end{figure}

For the purpose of this investigation, which was to test the performance of NN in identifying FMR spectra, we focused on two variables: the uniaxial anisotropy field $B_\mathrm{u}$ and the individual linewidth $\Delta B$. Therefore, for the training of the NN, we computed FMR spectra for a wide range of $B_\mathrm{u}$ and different values of $\Delta B$. Figure \ref{contour} shows examples of the training data used with $-250$ mT $\leq B_\mathrm{u}\leq +250$ mT and two different linewidths $\Delta B=20$ mT (Fig. \ref{contour}a) and $\Delta B=40$ mT (Fig. \ref{contour}b). From the contour plot, where each horizontal line is the projected FMR spectrum corresponding to each $B_\mathrm{u}$, the characteristic splitting of the resonances can be seen. The main resonance field $B_\mathrm{res}$ can be traced by the central white ridge, as indicated by arrows in \ref{contour}(a). 


\begin{figure}[h]
	\centering
		\includegraphics[width=1.00\columnwidth]{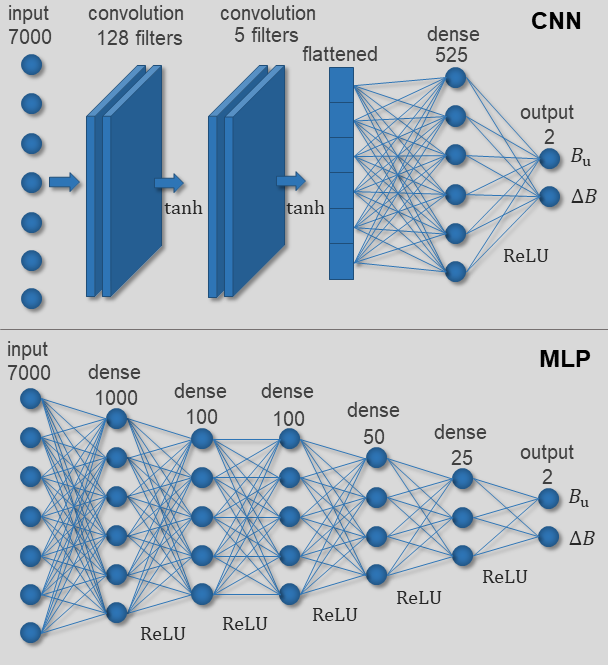}
	\caption{Neural Network architecture that goes from the FMR spectrum (input) to the system parameters (output) for the CNN (top) and the MLP (bottom), showing the number of neurons at each stage of the process and the activation function used at each step. The number of neurons is not to scale.}
	\label{NN_Architecture}
\end{figure}

\section{Neural Networks}

A Neural Network is a collection of connected units, called neurons, which maps inputs to outputs using connection weights and activation functions. Here, the input is the data points of the FMR spectrum and the output is the material parameters $B_\mathrm{u}$ and $\Delta B$. Neurons are typically arranged in layers and each layer may or may not be connected directly to any other layer. Information propagates from the first layer (the input layer) to the last layer (the output layer) through intermediate layers which are called hidden layers. We considered two types of networks: the Convolutional Neural Network (CNN) and the Multi-Layer Perceptron (MLP). Both are feed-forward networks, i.e., information only propagates in one direction. The difference between the two architectures is the types of hidden layers they employ: CNN uses at least one convolution layer, whereas MLP uses only dense layers (see Fig. \ref{NN_Architecture}). A convolution layer has a fixed-length window that moves along the inputs applying a filter at some step size, called stride (see animation in Suppl. Mat.). A filter is a subset of the connection weights that are multiplied to the inputs in the window, and this information is fed to a neuron in the convolution layer. A convolution layer can have multiple filters, and a particular filter will pick out a particular feature of the data. In contrast, in a dense layer each input from the previous layer is connected to every neuron. For the propagation of information from one layer to the next, the values of each neuron are multiplied by their associated weights and summed, an activation function is applied on that sum, and the information is passed on to the next layer. Commonly used activation functions are sigmoid, hyperbolic tangent and rectified linear activation function (ReLU), and as shown in Fig. \ref{NN_Architecture}, we found that the best choice for this work was $\tanh (x)$ and ReLU for the CNN and ReLU for the MLP.

Figure \ref{NN_Architecture} shows the details of the architecture for the neural networks used in this work. The CNN consisted of the input layer, two convolution layers, a flattened layer, a dense (fully connected) layer and the output layer, which is also a dense layer but without an activation function. The first convolution layer had 128 filters, a window size of 70 and a stride of 5, and it used $\tanh(x)$ as the activation function. The second convolution layer had 5 filters, a window size of 10 and a stride of 1, and it also used $\tanh(x)$ as the activation function. The next layer was used to flatten the outputs of the second convolution layer, i.e., to concatenate the filters into one row. Finally, the hidden dense layer had 525 neurons which were fully connected to the inputs from the flattened layer and to the outputs, and ReLU was used as activation function. The MLP consisted of the input layer, 5 hidden dense layers which all use ReLU activation, and the output layer. The size of the dense layer was gradually decreased from containing 1000 neurons to containing 25 neurons. For both NN, the output had 2 neurons: $B_\mathrm{u}$ and $\Delta B$.

Both NN were trained by supervised learning: each training sample (FMR spectrum) was provided with its desired output ($B_\mathrm{u}$ and $\Delta B$), and the training was performed by adjusting the weights in order to improve accuracy, i.e., how well the predicted values agree with the target values. During each training cycle, called an epoch, the weights were adjusted by means of a stochastic gradient-descent optimization algorithm (\textit{Adam}) \cite{Adam}, and the computations were implemented using \textit{Tensorflow} \cite{tensorflow2015}. 
The training was performed for 60 epochs for both NN, in order to compare their training evolution (see Fig. \ref{training}a,d). The agreement between target and prediction was quantified with the loss function, which was calculated with
\begin{equation}
L=\frac{1}{N}\sum_i^N\left\{\log \left[\cosh\left(B_{\mathrm{u},i}^p-B_{\mathrm{u},i}^t\right)\right]+\log\Big[\cosh\left(\Delta B_i^p-\Delta B_i^t\right)\Big]\right\}
\end{equation}
where the superscript $p$ and $t$ is for predicted and target value of the $i$-th data set, respectively. The logcosh function was chosen because it enables better learning, and it is stable in gradient-descent-based weight adjustment \cite{moshagen2021}.

\section{Results and Discussion}

\begin{figure}[h]
	\centering
		\includegraphics[width=1.00\columnwidth]{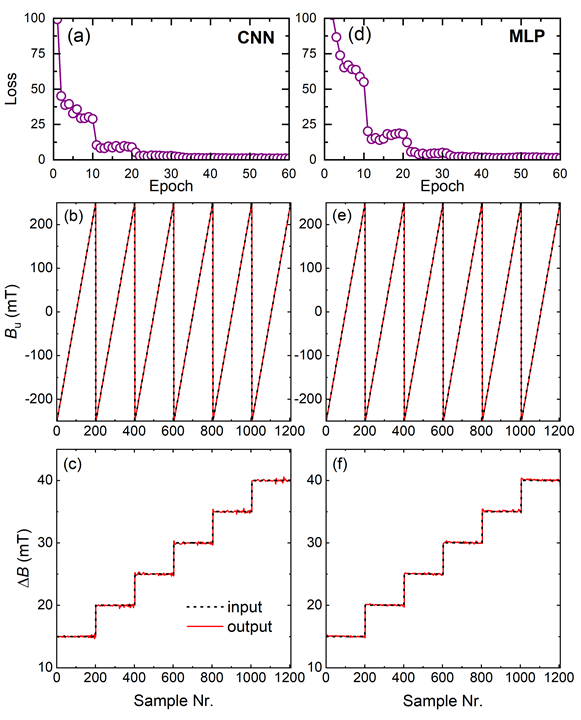}
	\caption{NN training: Panels (a) and (d) show the loss function with increasing number of epochs. For both NN the loss function converges rapidly and reaches unity after 60 epochs. Panels (b) and (e) show the training of the CNN and MLP with the anisotropy field $B_\mathrm{u}$, respectively, and panels (c) and (f) show the corresponding training with the linewidth $\Delta B$. Target values are shown as a dashed line and predicted values are shown as a solid red line. }
	\label{training}
\end{figure}

Figure \ref{training} presents an overview of the training process for each NN in the form of the output parameters as a function of sample number, where each sample number corresponds to a different set of $B_\mathrm{u}$ and $\Delta B$. For both NN, the loss function exhibits a steep decrease during the first cycles of training, followed by a plateau and then another steep decrease after 10 epochs and then gradually reaches unity within 60 epochs, as seen in Fig. \ref{training}(a) for the CNN and in Fig. \ref{training}(d) for MLP. 

Figure \ref{training} panels a--c show data for the CNN and panels d--e the MLP network, and they reveal the successful training of the networks. This can be seen by the excellent agreement between the target values for $B_\mathrm{u}$ and $\Delta B$, shown as a dashed line, and the predicted values, shown as a solid red line. The NN training resulted to a convergence between prediction and target for the entire training range with 201 values for $B_\mathrm{u}$ and 6 values for $\Delta B$.

\begin{figure}[h]
	\centering
		\includegraphics[width=1.00\columnwidth]{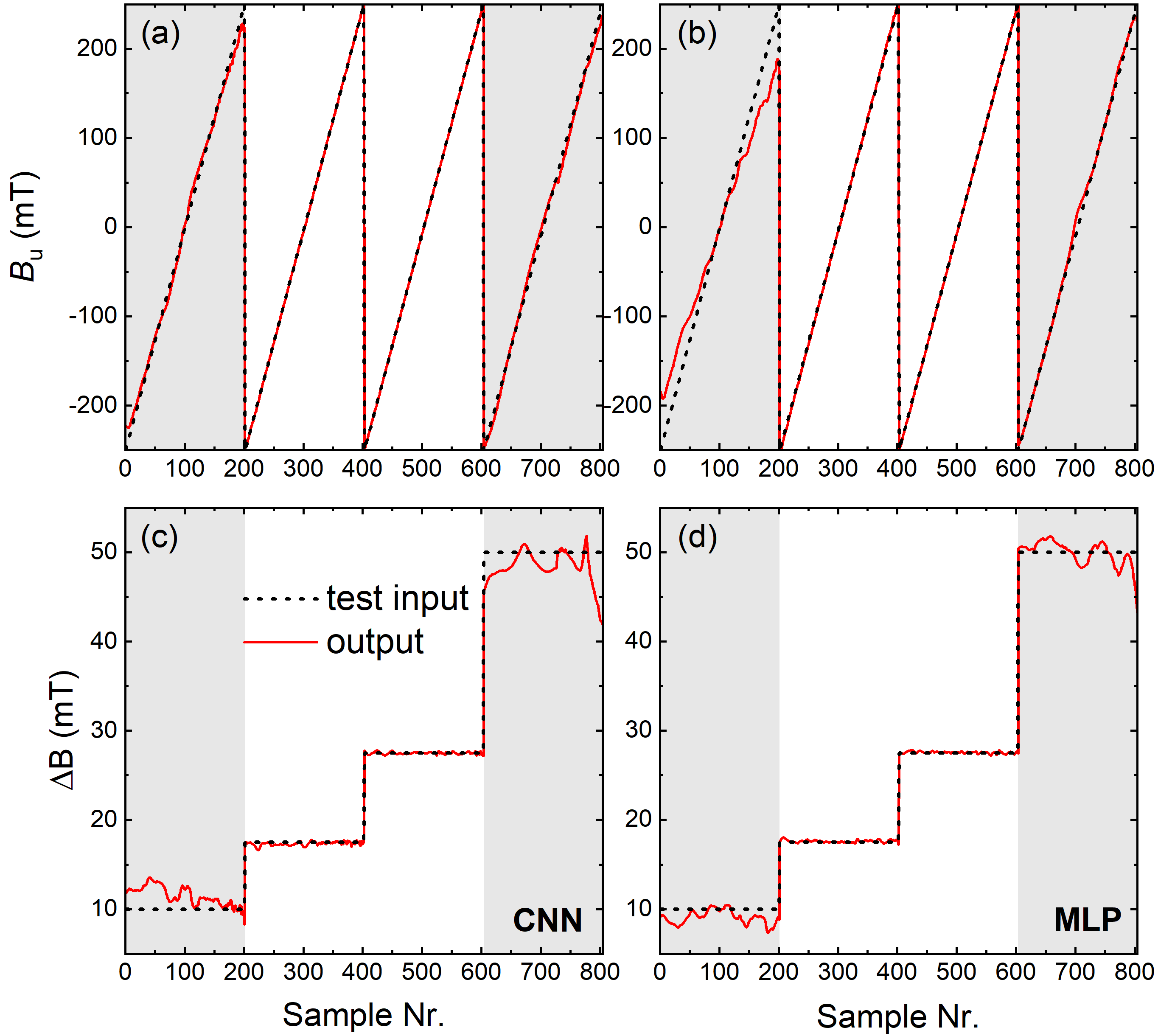}
	\caption{Testing of the NN: FMR data computed with anisotropy field in the range $-250$ mT $\leq B_\mathrm{u}\leq +250$ mT and four different linewidths $\Delta B=10$,17.5, 27.5, and 50 mT, were used as test-inputs. Both NN perform excellent when the input data is within the range of the training data (sample Nrs. 202 -- 603), but they also perform well when the input data is not within the training range (shaded regions). }
	\label{testing}
\end{figure}

Once the training of the NN is complete, the weights are fixed. An FMR spectrum that is introduced to the NN will undergo the transformations based on the individual architecture and the NN will produce a prediction for the parameters $B_\mathrm{u}$ and $\Delta B$. Both NN were tested with unseen FMR spectra, i.e., 804 spectra outside of the training data generated with the same 201 values for $B_\mathrm{u}$ and four values for $\Delta B$ (10, 17.5, 27.5 and 50 mT). Part of the test data fell within the range of the training set ($\Delta B=17.5$ mT and $27.5$ mT) and part of it was beyond the training set ($\Delta B=10$ mT and $50$ mT). In Fig. \ref{testing}, the shaded regions correspond to data that lies outside of the training range, and while the agreement between prediction and target is not as good as for the data that was within the training range, it is remarkable that both NN were able to come close to the truth and predict material parameters that were in reasonable agreement to the real values.

\begin{figure}[h]
	\centering
		\includegraphics[width=1.00\columnwidth]{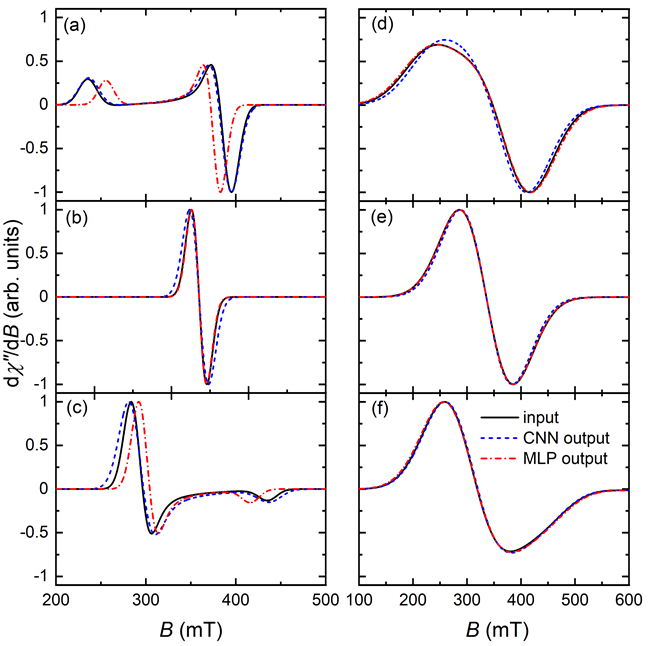}
	\caption{Comparison between selected FMR spectra showing spectra computed using the target values (solid line) and spectra computed using the predicted values of CNN (dashed line) and MLP (dashed-dotted line) for three anisotropy fields: (a) \& (d) $B_\mathrm{u}=+250$ mT; (b) \& (e) $B_\mathrm{u}=0$; (c) \& (f) $B_\mathrm{u}=-250$ mT for two different linewidths $\Delta B=10$ mT (left columns) and $\Delta B=50$ mT (right columns). These spectra correspond to Sample Nrs. 201 (a), 101 (b), 1 (c), 804 (d), 704 (e), and 604 (f) from Fig. \ref{testing}, and (a) represents the worst prediction, i.e., maximum deviation between target and predicted values of the entire data set.  }
	\label{fmr-spectra}
\end{figure}

In comparing the two NN, both performed exceptionally well for the data that was within the training range, whereas for data outside the training range, the CNN performed better in predicting $B_\mathrm{u}$ and the MLP performed better in predicting $\Delta B$. The performance of both NN can be assessed from the agreement between predicted and target values for both $B_\mathrm{u}$ and $\Delta B$, which we quantified with the Mean Absolute Error (MAE), shown in Table \ref{MAE}. The MAE for the training and the test with in-range data is comparable for both NN and it is on average smaller than 0.5 mT. For out-of-range data, the MAE is larger but still on the order of a few mT. This suggests that a properly trained NN can identify the correct material parameters from an FMR spectrum with relatively high accuracy.

\begin{table}
	\centering
		\begin{tabular}{|c|c|c|c|}
		\hline
		MAE $B_\mathrm{u}$ (mT) & Training & Test in range & Test out of range \\ \hline		
		CNN & $0.218$ & $0.365$ & $6.749$ \\ \hline
		MLP & $0.378$ & $1.215$ & $16.02$ \\ \hline \hline
		MAE $\Delta B$ (mT) & Training & Test in-range & Test out of range \\ \hline		
		CNN & $0.080$ & $0.155$ & $1.707$ \\ \hline
		MLP & $0.092$ & $0.119$ & $1.036$ \\ \hline  
		\end{tabular}
	\caption{Mean Absolute Error of the uniaxial anisotropy field $B_\mathrm{u}$ and the internal linewidth $\Delta B$ for the two NN. The MAE for the training and the test with in-range data is comparable, whereas the MAE for out-of-range data is an order of magnitude larger. The CNN performs better in finding $B_\mathrm{u}$ and the MLP performs better in finding $\Delta B$.}
	\label{MAE}
\end{table}

Figure \ref{fmr-spectra} shows selected examples of FMR spectra from the test set, which were outside the $\Delta B$ training range versus the associated \emph{predicted} FMR spectra computed using parameters predicted by the NN. It appears that for smaller linewidth the deviations are larger than for broader spectra. Despite this and the fact that the MAE is larger for out-of-range data, it is remarkable that both NN were able to predict reasonable $B_\mathrm{u}$ and $\Delta B$ outside their training range. Considering the simple model we investigated here, with only two parameters, $B_\mathrm{u}$ is a material parameter and does not depend on the measurement technique, thus CNN appears to be a better option for analyzing FMR spectra because it exhibits better performance in predicting $B_\mathrm{u}$.

Given the above, the use of neural networks for the spectral regression in order to obtain the material parameters is quite promising and further investigations are needed to fully harvest the potential of these techniques. Important future steps are to investigate FMR spectra with more parameters, i.e., to introduce additional anisotropy fields typically present in real materials, such as cubic anisotropy fields or surface anisotropy fields. Additionally, these techniques could be enhanced by tailoring the loss function to the particular problem, in what is termed as physics-informed machine learning.

\section{Conclusions}
We have shown here that artificial neural networks, specifically with the examples of a convolutional neural network and a multi-layer perceptron, inverse problems of ferromagnetic resonance can be solved by training the network with pre-computed FMR spectra and creating associations between material parameters and spectrum structure. Both architectures were able to successfully predict the correct material parameters by identifying the FMR spectra with an absolute mean error smaller than 1 mT. Strikingly, the networks were able to predict material parameters that were not part of the training set with a reasonable agreement to the real values, showing potential for extrapolation within a reasonable parameter range. These results highlight the potential of using neural networks in spectroscopic data analysis for the identification of material properties, and with further optimization of the architecture and training protocols, these techniques can be used for high-throughput spectroscopic analyses of magnetic nanoparticles for a wide range of applications.

\section*{ACKNOWLEDGMENTS}
MC is grateful for funding by the Louisiana Board of Regents [Contract No. LEQSF(2020-23)-RD-A-32]. Portions of this research were conducted with high performance computational resources provided by the Louisiana Optical Network Infrastructure (\href{http://www.loni.org}{http://www.loni.org}).

\section*{Data Availability}
The data that support the findings of this study are available from the corresponding author upon reasonable request.


%

\end{document}